\documentclass[prb,aps,showpacs,floatfix,floats,nobalancelastpage,twocolumn]{revtex4-1}
\usepackage{graphicx}
\usepackage{graphics}
\usepackage{latexsym,amsmath,amssymb,bm,euscript}
\usepackage{color}
\usepackage{dcolumn}
\usepackage{bm}
\usepackage{epstopdf}
\usepackage{times}
\usepackage{multirow}
\usepackage{wasysym}
\usepackage{subfigure}
\usepackage{bbm}
\def\etal{~\textit{et~al.}}
\def\ra{\rangle}
\def\la{\langle}

\def\Hc{{\rm H.c.}}

\def\linepic{{\begin{picture}(12, 0)(0,0)
                    \put(0,0) {\circle*{5}}
                    \put(12,0) {\circle*{5}}
                    \put(0,0){\line (1,0) {12}}
                  \end{picture}}}

\def\triangpic{{\begin{picture}(17,15)(-2,-2)
                      \put (0,0) {\circle*{5}}
		      \put (12,0) {\circle*{5}}
		      \put (6,10) {\circle*{5}}
		      \put (0,0) {\line (1,0) {12}}
		      \put (12,0) {\line (-3,5) {6}}
		      \put (0,0) {\line (3,5) {6}}
                \end{picture}}}

\begin{document}

\title{Possible uniform-flux chiral spin liquid states in the SU(3) ring-exchange model on the triangular lattice}
\author{Hsin-Hua Lai}
\affiliation{National High Magnetic Field Laboratory, Florida State University, Tallahassee, Florida 32310, USA}
\date{\today}

\begin{abstract}
We consider a SU(3) model with antiferromagnetic three-site ring exchanges, in addition to two-site exchanges, on the triangular lattice. We first present numerical site-factorized state studies on the magnetic ordered states, which shows two different three-sublattice-ordered states, the antiferro-quadrupolar phase and the standard $120^o$ anti-ferromagnetic phase,  along the axis of the strength of the three-site ring exchanges. We further study the model using slave-fermion mean field approaches in which we rewrite the exchange operators in terms of three flavors of fermions. At the mean-field level, we find the main competing trial states are the trimer state (triangular plaquette state), gapped uniform $\pi/3$-flux chiral spin liquid, and gapped uniform $2\pi/3$-flux chiral spin liquid. The filled band of the $\pi/3$-flux chiral spin liquid has Chern number $+1$, and that of the $2\pi/3$-flux chiral spin liquid has Chern number $+2$. We also give the effective Chern-Simons theory for each chiral spin liquid at the mean-field level.
\end{abstract}
\maketitle

\section{Introduction}\label{Introduction}
One of the goals in the studies of strongly interacting cold atom gases is that they can be used to simulate strongly correlated systems \cite{Bloch_RMP}. The model systems can be engineered with a high degree of control and experimentally studied to reveal the profound nature of the phases, among which the Quantum spin liquid states (QSL) \cite{LeeNagaosaWen,Balents_nature} are perhaps one of the most intriguing phases. With cold atom gases, the $N$-flavor fermionic Hubbard model on different lattices can be also realized.  A model Hamiltonian to describe such systems is the $N$-flavor fermionic Hubbard model \cite{Honerkamp2004, Xu2010, Gorshkov10}
\begin{eqnarray}
\nonumber H = - t \sum_{\la jk \ra}\sum_{\alpha} \left[ c^{\alpha\dagger}_j c^{\alpha}_k + \Hc \right] + U \sum_{j}\sum_{\alpha,\beta}n^{\alpha}_j n^{\beta}_j,~
\end{eqnarray}
where $\alpha,~\beta$ run over the different flavors, $\la jk \ra$ runs over pairs of nearest neighbors on the lattice, and $j$ runs over all lattice sites.

If we focus on the $1/N$ filling, the system with generic $N$ flavor of fermions can undergo metal-to-Mott insulator phase transition for sufficiently large repulsion $U$. The transition to a Mott insulator has been recently observed in ($N=2$) spin-$1/2$ Hubbard model. \cite{Robert08, Schneider2008} In this case, it is generally accepted that the ground state is well-captured by the usual antiferromagnetic (AFM) Heisenberg model. On the other hand, when $U$ is not very large compared with the hopping strength $t$, the ground state is not well-understood and it is possible that the strong charge fluctuations near the transition play an important role for stabilizing ``gapless'' spin liquid phases. \cite{ringxch,LeeandLee05,HYYang2010}

For $N > 2$, \cite{Affleck1988, Read1989, Read1990, Harada2003, Naoki2007, Beach2009, Hermele2009, Hermele2011, Cai2012} in the large $U$ limit we also obtain Heisenberg-like (two-site exchange) Hamiltonian
\begin{eqnarray}
\nonumber H_{2} = J\sum_{\la jk \ra} P_{jk} ,~
\end{eqnarray}
where $P_{jk}$ is the two-site exchange operator, which permutes the fermions between two nearest-neighbor sites as $P_{jk} | \alpha,~ \beta \ra = | \beta,~ \alpha \ra$, where the $\alpha,~\beta$ represent the spin states at sites $j$ and $k$. For $N =3$, there has been numerical evidence on such a SU(3) Heisenberg model on the triangular lattice suggesting three-sublattice-ordered ground state in this regime. \cite{Bauer2012} 

However, at $t/U\sim O(1)$ in the SU(3) case, the two-site exchange Hamiltonian is not sufficient to capture the essential physics and higher ordered contributions should be considered. If we perform perturbation studies on the SU(3) Hubbard model at $1/3$-filling, we will in fact obtain ''ferromagnetic'' (FM) three-site ring exchanges and four-site ring exchanges and etc., if higher-order terms are included. In the FM three-site ring exchanges regime with four-site ring exchanges included, in Ref.~\onlinecite{Lai_SU(3)gaplessSL} we explored the magnetic ordered phases using the site-factorized ansatz, which found three-sublattice ordered states and FM state in a large parameter regime;we also explored the non-magnetic ordered phases using the mean-field slave-fermion trial states. By qualitatively arguing that the quantum fluctuations in linear flavor wave theory approximations destroy the three-sublattice ordered states in strong ring exchanges, We conjectured that several possible {\it gapless} QSL were present due to the geometrical frustration along with the strong ring exchanges near the metal-Mott phase transition. 

For the AFM side of three-site ring exchange regime, at the first sight it seems unnatural to consider this regime. However, from the perspective of cold atom systems, recently the cold atom experiment by Struck \etal~\cite{Struck2012} demonstrated a method to be able to add an artificial tunable gauge potential to the system. With the tunable gauge potential, it is possible to tune the sign of the three-site ring exchanges from FM to AFM. In addition, recently, Ref.~\onlinecite{Bieri2012} did variational studies on the SU(3) model with three-site ring exchange (no higher-order terms such as four-site ring exchanges and etc.), and found, on the AFM side of the three-site ring exchanges, an interesting $d_x + i d_y$ chiral spin liquid state (CSL) with a gapless parton Fermi surface. However, the uniform-flux  chiral spin liquid states, \cite{Kalmeyer1987,Kalmeyer1989,Laughlin1990,WenWilczekZee_CSL, Yang1993} QSL that break both parity and time-reversal symmetries,  were not considered in the previous studies. In this paper, we focus on the SU(3) model with ``AFM`` three-site ring-exchange on the triangular lattice and we find, at the mean field level, the energies of uniform-flux CSL are lower than the interesting $d_x + i d_y$ CSL and suggest that they are stabilized in this model, at least at the mean field.

In order to obtain all possible ground states including magnetic ordered and non-magnetic phases, in this work, we first study the magnetic ordered phases using the site-factorized ansatz.\cite{Lauchli2006} We find that the phase diagram contains two different three-sublattice-ordered phases, the same to those found in Ref.~\onlinecite{Bauer2012}. We further study this model using mean-field slave-fermion trial states focusing on the non-magnetic phases. After performing numerical full optimization of the trial energy, we consider three ansatz states--the uniform $\pi/3$-flux chiral spin liquid state ($\Phi_{\frac{\pi}{3}}$ CSL), the uniform $2\pi/3$-flux chiral spin liquid state ($\Phi_{\frac{2\pi}{3}}$ CSL), and the trimer (plaquette) state. At the mean-field level, each gapped uniform-flux state breaks Time Reversal (TR) and possesses a finite Chern number, $C=+1$ for the $\Phi_{\frac{\pi}{3}}$ CSL and $C=+2$ for the $\Phi_{\frac{2\pi}{3}}$ CSL.

The paper is organized as follows. In Sec.~\ref{Sec:model} we define explicitly the model Hamiltonian we will study. In Sec.~\ref{Subsec:site_factor} we use the site-factorized ansatz to study the magnetic ordered states. In Sec.~\ref{Subsec:MFcalc} we use the slave-fermion representation to rephrase the SU(3) Hamiltonian in terms of the three-flavor fermionic Hamiltonian and perform the fermionic mean-field treatment of the model. In Sec.~\ref{Subsec:Chern number} we give the effective Chern-Simons theories for each CSL stabilized in this SU(3) ring exchange model at the mean-field level. In Sec.~\ref{Sec:Discussion} we conclude with some discussions.
\section{SU(3) Model with ring exchange terms}\label{Sec:model}
\begin{figure}[t]
\includegraphics[width=\columnwidth]{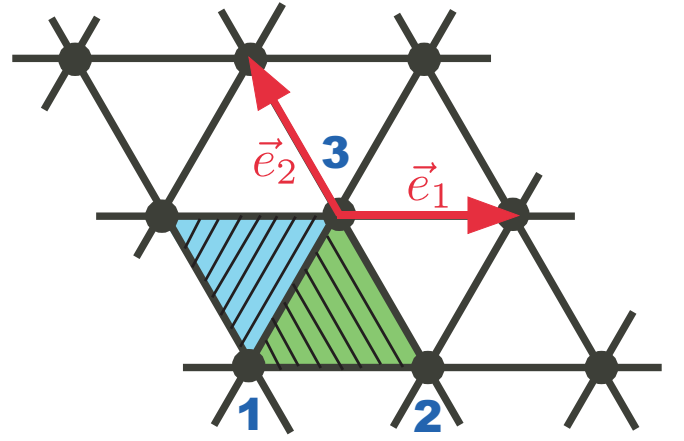}
\caption{The illustration of the triangular lattice. For each triangle, we label each site counterclockwise from 1 to 3. For each site, there are two triangles, a green backslashed up triangle and a blue forwardslashed down triangle, associated with it.
}
\label{triangular_lattice}
\end{figure}

The model Hamiltonian we consider is
\begin{eqnarray}\label{SU(3)_H}
H_{SU(3)} = && J \sum_{\linepic} P_{12} + K_{3} \sum_{\triangpic}  \left[ P_{123} +\Hc \right],
\end{eqnarray}
with $~\linepic~$ running over all the bonds on the lattice;$~\triangpic~$ running over all the triangles, up- and down-triangles, on the lattice, and $\la 123 \ra$ are the sites on the triangles labeled counterclockwise, Fig.~\ref{triangular_lattice}. $P_{12}$ is the nearest-neighbor two-site exchange operator and $P_{123}$ is the three-site ring exchange operator, which permutes the fermions on the triangles as $P_{jkl} |\alpha, \beta, \gamma \ra = |\gamma, \alpha, \beta \ra$. Without the ring exchange terms, previous studies of the site-factorized ansatz on the triangular lattice predicted a three-sublattice-ordered state \cite{Papanicolaou1988, Tsunetsugu2006, Lauchli2006} which was recently confirmed by Density Matrix Renormalization Group (DMRG) and infinite Projected Entangled-Pair States (iPEPS) analysis \cite{Bauer2012}. 

In this section, we will focus on the SU(3) model with AFM three-site ring exchanges. In order to study the magnetic ordered phases, we first study the model using the site-factorized states below in Sec.~\ref{Subsec:site_factor}. Later in Sec.~\ref{Subsec:MFcalc} we will present our studies using the slave-fermion trial states focusing on the non-magnetic phases.
\subsection{Site-factorized state studies}\label{Subsec:site_factor}

\begin{figure}[t]
\includegraphics[width=\columnwidth]{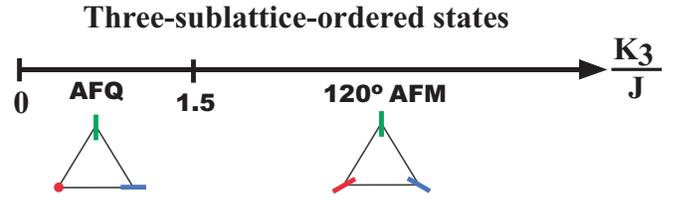}
\caption{
The phase diagram using the site-factorized states. There are two different three-sublattice-ordered states. One of them is the antiferro-quadrupolar phase (AFQ) whose on-site vector, Eq.~(\ref{site-factorized-vect}), is mutually orthogonal to each other, and the other is the standard $120^o$ AFM phase which can be characterized by calculating the inner products between each pair of the different nearest-neighbor on-site vectors.
}
\label{site_factorized_pdg}
\end{figure}

In this subsection, we consider the site-factorized state \cite{Lauchli2006} defined as
\begin{eqnarray}\label{site-factorized-vect}
|s \ra = \prod_{j} | \mathcal{X}_j \ra,
\end{eqnarray}
with
\begin{eqnarray}\label{site_factorized_study:parameterization}
| \mathcal{X}_j \ra \equiv a_{j} | x \ra_{j} + b_{j}  | y\ra_{j} + c_j  | z \ra_{j},
\end{eqnarray}
where we fix the overall phase by setting the phase of $a_j$ to be zero such that $a_j \in~\mathbb{R}$ and $b_j,~c_j~\in~\mathbb{C}$ and $|a_j|^2 + |b_j|^2 + |c_j|^2 =1$. Above, we used the usual time-reversal invariant basis of the SU(3) fundamental representation \cite{Lauchli2006}, defined as
\begin{equation}\label{xyz_transf}
\begin{array}{ccc}
|x \ra = \frac{i | 1 \ra - i | -1 \ra}{\sqrt{2}}; & |y\ra =\frac{|1 \ra + | -1 \ra}{\sqrt{2}}; & |z \ra =-i | 0 \ra,
\end{array}
\end{equation}
with $|S^z = \pm 1 \ra \equiv |\pm1\ra$ and $|S^z = 0\ra \equiv |0\ra$.

According to the parametrization of the $|\mathcal{X}_j\ra$ vector in Eq.~(\ref{site_factorized_study:parameterization}) along with the constraint, at each site there are $4$ independent parameters. For a lattice with $N\times N$ sites, there are $4N^2$ independent parameters for the site-factorized state, Eq.~(\ref{site-factorized-vect}). We numerically find the optimized (lowest) site-factorized state energy, $E_{sf} = \la s| H_{SU(3)} |s \ra$, on a $3\times3$, and on a $6\times6$ triangular lattice for a certain $K_3$ while $J\equiv 1$ using the gradient descent method.

The phase diagram is shown in Fig.~\ref{site_factorized_pdg}. Compared with the site-factorized state studies on the SU(3) model with the FM $K_3$, which contains the FM ordered state in addition to the three-sublattice-ordered phase, called the antiferro-quadrupolar phase (AFQ) in this paper \cite{Lai_SU(3)gaplessSL}, the phase diagram along the AFM $K_3/J$-axis contains two different three-sublattice-ordered states, the AFQ and the standard $120^o$ AFM. The AFQ is the same to the three-sublattice-ordered phase found in the FM $K_3$ side, \cite{Lai_SU(3)gaplessSL} with each on-site vector ($|\mathcal{X}_{j\in A}\ra, |\mathcal{X}_{j \in B}\ra,$ and $|\mathcal{X}_{j \in C}\ra$) mutually orthogonal to each other. The $120^o$ AFM can be characterized by calculating the inner products between each pair of the on-site vectors.

The energies of both states can be calculated analytically. The site-factorized state energy of the AFQ is exactly zero ( $E_{AFQ} =0$) and that of the AFM is $E_{AFM} =\frac{3}{4}J -\frac{1}{2} K_3$. The transition between these two states is at $K_3/J = 3/2$, which is consistent with our numerical calculation. We emphasize that site-factorized ansatz studies though give poor energies due to the ignorance of the quantum fluctuations, it can give us the correct order in energetics among the magnetic ordered phases. Comparing the result of magnetic ordered phases with that in Ref.~\onlinecite{Bieri2012}, we can see the order in energetics between only the magnetic ordered phases are consistent. For the corrections for the energies of these magnetic ordered phases, we can perform linear flavor wave theory calculations but these calculations still can not allow us to compare the energetics between the magnetic ordered states here and the non-magnetic ordered states obtained in the next subsection.
\subsection{Slave-fermion trial states and energetics}\label{Subsec:MFcalc}
In this subsection, we follow the approach similar to the one outlined in Ref.~\onlinecite{Serbyn2011} for the $S = 1$ spin model. We write the spin operators in terms of three flavors of fermionic spinons, $f^{\alpha}$,
\begin{eqnarray}
S^{\alpha}_j = - i \sum_{\beta, \gamma} \epsilon^{\alpha \beta \gamma} f^{\beta \dagger}_j f^{\gamma}_{j},~
\end{eqnarray}
with $\alpha,~\beta,~\gamma \in \{x, y, z\}$ and $j$ is the site label. Rewriting the spin operator in terms of fermionic spinons enlarges the Hilbert space. To recover the physical subspace, a local constraint on the fermions has to be enforced,
\begin{eqnarray}\label{constraint}
\sum_{\alpha}f^{\alpha \dagger}_{j} f^{\alpha}_{j} = 1.
\end{eqnarray}
The exchange operators in terms of fermions are
\begin{eqnarray}
&&P_{jk} = \sum_{\alpha \beta} f^{\alpha \dagger}_{j} f^{\beta}_{j} f^{\beta \dagger}_{k} f^{\alpha}_{k},\\
&& P_{jkl} = \sum_{\alpha \beta \gamma} f^{\alpha \dagger}_{j} f^{\beta}_{j} f^{\beta \dagger}_{k} f^{\gamma}_{k} f^{\gamma \dagger}_{l} f^{\alpha}_{l},
\end{eqnarray}
where $\sum_{\alpha} =\sum_{\alpha= x,~y,~z}$ and similar relations for $\beta$ and $\gamma$.

The Hamiltonian, Eq.~(\ref{SU(3)_H}), can be re-expressed as
\begin{eqnarray}
\nonumber H_{SU(3)} =&& J \sum_{\linepic} \sum_{\alpha \beta}  f^{\alpha \dagger}_{1} f^{\beta}_{1} f^{\beta \dagger}_{2} f^{\alpha}_{2} + \\
\nonumber  && \hspace{-0.5cm} + K_3 \sum_{\triangpic} \sum_{\alpha \beta \gamma} \bigg{[} f^{\alpha\dagger}_1 f^{\beta}_{1} f^{\beta \dagger}_{2} f^{\gamma}_2 f^{\gamma \dagger}_3 f^{\alpha}_{3} + \Hc \bigg{]}.~~~~~~
\end{eqnarray}
Below, we will calculate the trial energies using the slave fermion trial states. In order to check which mean-field ansatz states are most relevant for this model, we perform numerically ``full optimization'' of the mean-field energy, Eq.~(\ref{MF_energy:general_expression}), on a triangular lattice with $100 \times 100$ 3-site unit cells, by treating $\chi_{jk}$-s and $\theta_{jk}$-s as varying variables. In the numerical optimization, there are totally $18$ variables ($9$ $\chi_{jk}$ and $9$ $\theta_{jk}$) and we take $t_{jk} =1$, $\mu^{\alpha}_j = \mu$. Numerics suggest that, at the mean-field level, the main competing states are the ``trimer'' state, the $\Phi_{\frac{\pi}{3}}$ CSL, and the $\Phi_{\frac{2\pi}{3}}$ CSL. \cite{energetics_note}

When we perform numerical calculations, we relax the constraint of the fermion number for each flavor to be
\begin{eqnarray}\label{relaxed_constraint}
\la f^{\alpha \dagger}_j f^{\alpha}_j \ra_{trial}= \frac{1}{3}.
\end{eqnarray}
A convenient formulation of the mean field is to consider a general SU(3)-rotation invariant trial Hamiltonian
\begin{eqnarray}\label{MF_H:nopairing}
\nonumber H_{trial} = && - \sum_{\la jk \ra}\sum_{\alpha} \left[ t_{jk}e^{- i \theta_{jk}} f^{\alpha \dagger}_{j} f^{\alpha}_{k} + \Hc \right] -\\
&& -\sum_{j}\sum_{\alpha} \mu_{j}f^{\alpha \dagger}_{j} f^{\alpha}_j,~
\end{eqnarray}
with  $t_{jk}$ being the hopping amplitude, $\theta_{jk}$ being the phase of the hopping $t_{jk}$ in different mean-field ansatz states, and $\mu_j$ being the chemical potential which can be used to satisfy the constraint, Eq.~(\ref{relaxed_constraint}). With the trial Hamiltonian above, we can find the ground state and use it as a trial wave function for the Hamiltonian $H_{SU(3)}$, Eq.~(\ref{SU(3)_H}). After performing ``complete'' Wick contractions and ignoring the constant pure density terms, the trial energy can be expressed as
\begin{widetext}
\begin{eqnarray}\label{MF_energy:general_expression}
\nonumber E_{MF} = && -J \sum_{\linepic} \bigg{|} \sum_{\alpha} \chi^{\alpha}_{12} \bigg{|}^2 + \\
&& + K_3 \sum_{\triangpic} \bigg{\{}\bigg{[} \sum_{\alpha} \chi^{\alpha}_{12}  \chi^{\alpha}_{23} \chi^{\alpha}_{31}  - \sum_{\alpha \beta} \bigg{(} n^{\alpha}_{1} \chi^{\alpha}_{23} \chi^{\beta}_{32} +  n^{\alpha}_{2} \chi^{\alpha}_{31} \chi^{\beta}_{13} + n^{\alpha}_{3} \chi^{\alpha}_{12} \chi^{\beta}_{21} \bigg{)} +  \sum_{\alpha \beta \gamma} \chi^{\alpha}_{13} \chi^{\beta}_{32} \chi^{\gamma}_{21} \bigg{]} + \Hc \bigg{\}},
\end{eqnarray}
\end{widetext}
where we defined $( \chi_{jk}^\alpha)^* \equiv \la f^{\alpha \dagger}_j f^{\alpha}_k \ra_{trial}$.

\begin{figure}[t]
\subfigure[]{\label{CSL:pi/3} \includegraphics[width=\columnwidth]{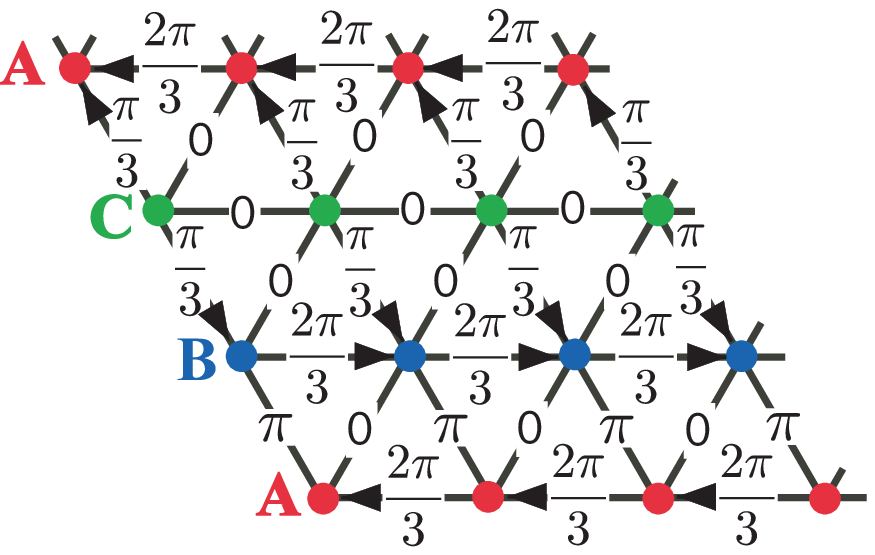}}
\subfigure[]{\label{CSL:2pi/3}\includegraphics[width=\columnwidth]{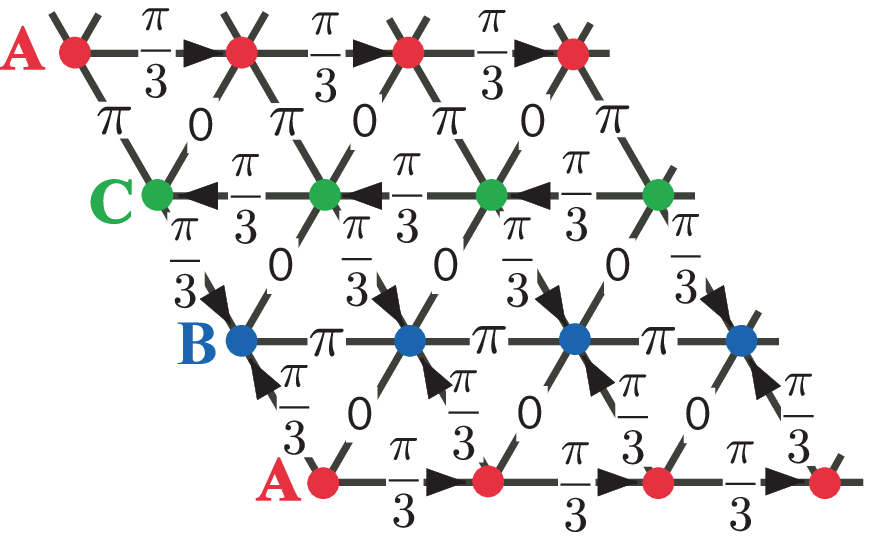}}
\caption{Graphical illustrations of the $\Phi_{\frac{\pi}{3}}$ CSL and $\Phi_{\frac{2\pi}{3}}$ CSL. Both of these states have three sublattices per unit cell labelled as A, B, and C. The directions of the arrows represent the phases associated with the fermion hoppings.The arrows represent the direction of the gauge. (a) $\Phi_{\frac{\pi}{3}}$ CSL. (b) $\Phi_{\frac{2\pi}{3}}$ CSL.}
\label{CSL:pic}
\end{figure}

The slave-fermion trial states which conserve the translational symmetry that we consider are the $\Phi_{\frac{\pi}{3}}$ CSL and $\Phi_{\frac{2\pi}{3}}$ CSL, Fig.~\ref{CSL:pic}. There are three sublattices per unit cell for $\Phi_{\frac{\pi}{3}}$ CSL and $\Phi_{\frac{2\pi}{3}}$ CSL and the directions of the arrows in Fig.~\ref{CSL:pic} represent the phases associated with the fermion hoppings. Below we list the numerical values of the energy per site in the two states
\begin{eqnarray}
&& E^{MF}_{\Phi=\pi/3} = -0.8992 J - 0.8343 K_3, \label{MF_pi/3:num_val}\\
&& E^{MF}_{\Phi=2\pi/3} = -0.5425 J - 0.8942 K_3 . \label{MF_2pi/3:num_val}
\end{eqnarray}

\begin{figure}[t]
\includegraphics[width=\columnwidth]{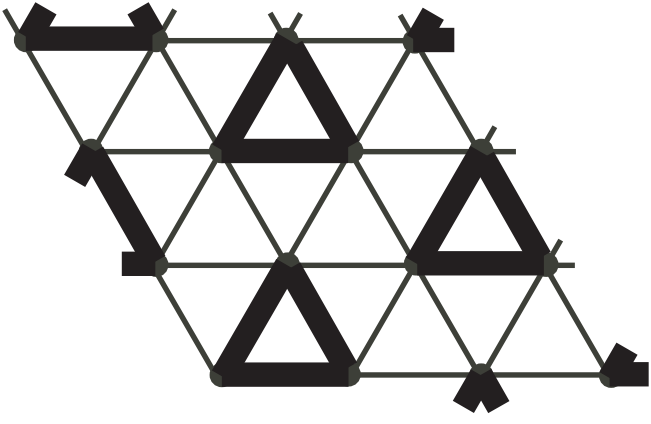}
\caption{Graphical illustration of the trimer state. In slave fermion picture, the fermionic spinons only hop around each triangular plaquette and we can focus on each triangle separately.}
\label{trimer}
\end{figure}

Besides the translationally invariant state, we also consider what we call the ``trimer'' state. Fig.~\ref{trimer} shows one example of the configuration of such a state in which the non-zero $t_{jk}$ form non-overlapping trimer covering of the lattice. These states break translational invariance, and any trimer covering produces such a state. Such states can have the lower Heisenberg exchange energy. The occupied bonds attain the maximal expectation value which is found analytically $|\chi^{\alpha}_{jk}|_{max} = n^{\alpha}_j = 1/3$. Their contribution can be sufficient to produce the lowest total energy and such states are expected to be the lowest-energy states with $K_3 = 0$.
\begin{eqnarray}
E^{MF}_{trimer} = - J - 0.5926K_3.~\label{MF_trimer:num_val}
\end{eqnarray}

\begin{figure}[t]
\includegraphics[width=\columnwidth]{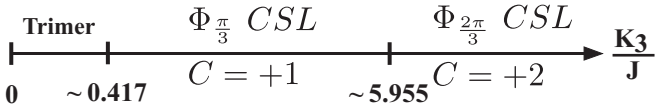}
\caption{The phase diagram of the mean-field ansatzes. The exact wave function of the trimer state can be written down explicitly, Eq.~(\ref{trimer:wf}), and we can calculate the corresponding energy exactly.
}
\label{MFSL_pdg}
\end{figure}

Based on the energies of each trial state, Eq.~(\ref{MF_pi/3:num_val})-(\ref{MF_trimer:num_val}), we can analyze the phase diagram along $K_3/J$-axis. At $K_3 =0$, the trimer state is the lowest energy state followed by the $\Phi_{\frac{\pi}{3}}$ CSL and the $\Phi_{\frac{2\pi}{3}}$ CSL. When $K_3$ increases, the energy line of the $\Phi_{\frac{\pi}{3}}$ CSL crosses that of the trimer state and $\Phi_{\frac{\pi}{3}}$ CSL becomes the lowest energy state at $K_3/J \sim 0.417$. When $K_3/J$ keeps increasing, the energy of the $\Phi_{\frac{2\pi}{3}}$ will be lower than that of the $\Phi_{\frac{\pi}{3}}$ CSL when $K_3 /J >5.955$ and becomes the lowest-energy state. The results are summarized in the mean-field phase diagram, Fig.~\ref{MFSL_pdg}. At the mean-field level, for each CSL, we also calculate the Chern number of the filled lowest band. The details will be illustrated in Sec.~\ref{Subsec:Chern number}. In short, at the mean-field level, the $\Phi_{\frac{\pi}{3}}$ CSL has a Chern number $C= +1$ and the $\Phi_{\frac{2\pi}{3}}$ CSL has a Chern number $C= +2$.

Before leaving this section, we want to remark that the trimer state is a singlet state around a triangular plaquette, and we can write down the exact singlet wave function  in a closed form as
\begin{eqnarray}\label{trimer:wf}
|\psi_{trimer}\ra = \sum_{\alpha, \beta, \gamma}\frac{\epsilon^{\alpha \beta \gamma}}{\sqrt{6}} | \alpha, \beta, \gamma \ra,
\end{eqnarray}
with $\alpha=x,y,z$. With the trimer wave function, we can calculate the exact energy per site
\begin{eqnarray}
\mathcal{E}_{\psi_{trimer}} = -\frac{1}{3} J +\frac{4}{27}K_3.
\end{eqnarray}
In general, the exact energy of the trimer is lower than that of the AFQ \cite{LFWT:AFQ_note} and is lower than that of the AFM when $K_3/J <117/70 \sim 1.67$. In order to compare the energies of the CSL  with other states, we need to perform Gutzwiller projections \cite{WenPSG} on the mean-field CSL states we obtain above, which is beyond the scope of this paper.
\subsection{Berry curvature and Chern number}\label{Subsec:Chern number}
\begin{figure}[t]
\subfigure[]{\label{Edge:pi/3} \includegraphics[width=\columnwidth]{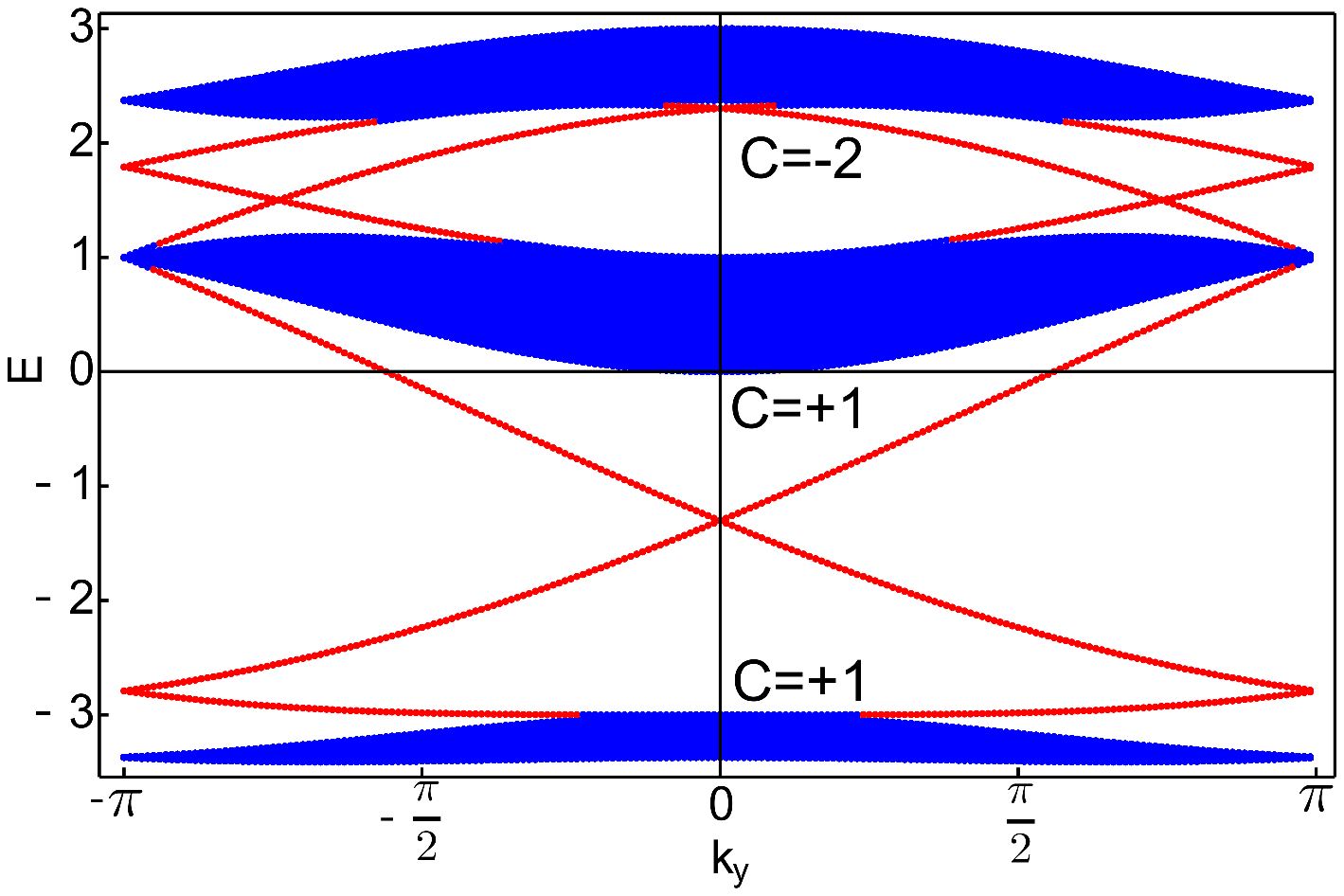}}
\subfigure[]{\label{Edge:2pi/3}\includegraphics[width=\columnwidth]{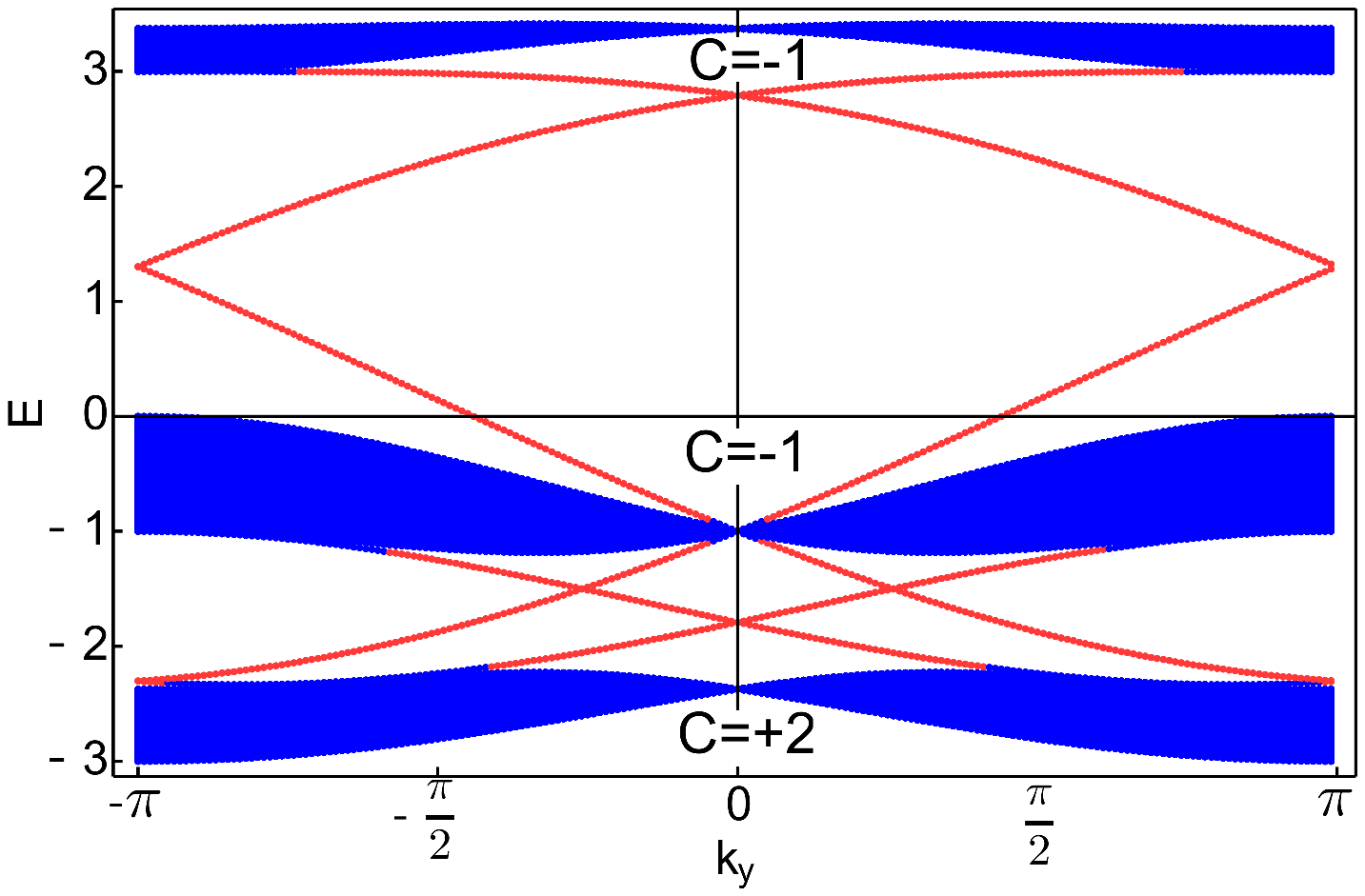}}
\caption{Edge states and Chern numbers of each band for a flavor of fermions in the trial uniform-flux states. The lowest band is completely filled for each flavor of fermions. (a) The edge states of the $\Phi_{\frac{\pi}{3}}$ CSL. Each flavor of fermions show the same edge states and bulk bands whose filled lowest band is of Chern number $C=+1$. (b) The edge states of the $\Phi_{\frac{2\pi}{3}}$ CSL. The lowest filled bands for three flavors of fermions show Chern numbers $C=+2$. The $\Phi_{\frac{2\pi}{3}}$ CSL can be obtained by inserting a $\pi$ flux to each triangle along with flipping the arrow directions. The two spectra can be related by $\epsilon_{\Phi_{2\pi/3}}({\bf k}) = - \epsilon_{\Phi_{\pi/3}} ({\bf k} + \pi$). }
\label{Edge_states}
\end{figure}

At the mean-field level we can characterize each CSL discussed above by calculating its corresponding Chern number \cite{TKNN1982}. Full theory can have other excitations which we will ignore in this paper. Since each flavor of fermions have a filling $\nu=1/3$ per site and forms a band insulator, there is a well-defined Chern number. The Chern number (which is the Berry phase in units of $2\pi$) of a many-body state at band $a$ is an integral invariant in the boundary phase space,
\begin{eqnarray}\label{Chern_number:formula}
C_a=\frac{i}{2\pi}\int_{BZ} dk_1dk_2 \left[ \bigg{\la} \frac{\psi_a}{\partial k_1}\bigg{|}\frac{\psi_a}{\partial k_2}\bigg{\ra} - \bigg{\la} \frac{\psi_a}{\partial k_2} \bigg{|} \frac{\psi_a}{\partial k_1}\bigg{\ra}\right],~~~~~
\end{eqnarray}
where $\psi_a$ is the wave-function of the band $a$, $k_{1/2}$ represent the momentum along vectors $\vec{e}_1$ and $\vec{e}_2$, Fig.~\ref{triangular_lattice}, and BZ stands for the first Brillouin zone. Using Eq.~(\ref{Chern_number:formula}), we obtain the Chern numbers for the lowest filled band in $\Phi_{\frac{\pi}{3}}$ CSL and $\Phi_{\frac{2\pi}{3}}$ CSL. We find that $C=+1$ for $\Phi_{\frac{\pi}{3}}$ CSL and $C=+2$ for $\Phi_{\frac{2\pi}{3}}$ CSL. Since the band spectra for each flavor of fermions are the same, we focus on one flavor of fermions. Fig.~\ref{Edge_states} shows the edge states of each CSL and the Chern numbers of each bulk band.

It is known that the effective theory of the bulks of CSL are described by the effective Chern-Simons theory, which gives the low-energy excitations with self- and mutual- braiding statistics. Besides, the sharpest differences between different CSL lie on the edges. According to the edge-bulk correspondence, \cite{Wen1995} we can obtain the corresponding edge theory from the Chern-Simons description of the bulks. The effective Chern-Simons theories hence are useful for characterizations of each CSL and will given below.

Before giving the effective Chern-Simons theory for the filled lowest fermion band in the CSL, it is convenient to transform the basis from $\{f^x, f^y, f^z\}$ to $\{f_{+1}, f_{-1}, f_{0}\}$, where $ f_{\pm 1}$ carry $S^z$ quantum number $\pm 1$ and $f_{0}$ carries $S^z$ quantum number $0$. Based on Eq.~(\ref{xyz_transf}), we can transform the fermions with
\begin{equation}
\begin{array}{ccc}
f^x = \frac{-i}{\sqrt{2}}\left[f_{+1} - f_{-1}\right],& f^y = \frac{1}{\sqrt{2}}\left[f_{+1} + f_{-1}\right], & f^z = i f_0.
\end{array}
\end{equation}
In $\{f_{\pm1}, f_{0}\}$ basis, the spin operator can be represented as
\begin{eqnarray}
&& S^{+} \equiv S^x + i S^y = \sqrt{2} \left( f^\dagger_{+1} f_0 + f^\dagger_0 f_{-1}\right),\\
&& S^z = f^\dagger_{+1} f_{+1} - f^\dagger_{-1} f_{-1}.
\end{eqnarray}
In this new basis, below we will give the effective Chern-Simons theory for each CSL and its corresponding edge theory at the mean field.
\subsubsection{Effective Chern-Simons theory for mean-field $\Phi_{\frac{\pi}{3}}$ CSL}\label{Subsubsec:CS4pi/3}
The three flavors of fermions, $\{f_{\pm1},~f_0\}$, fill the lowest bands with Chern numbers $C_{\pm1}=C_0=+1$. Conserved fermion currents $J^\mu_m$ can be expressed in terms of dynamical $U(1)$ gauge fields $a^m_{\mu}$ as $J^\mu_m = \frac{\epsilon^{\mu\nu\lambda}}{2\pi} \partial_\nu a^m_{\lambda}$ with $m=\pm1,~0$, where summation over repeated $\mu,\nu,\lambda$ is assumed. The fermion band structure in this $\Phi_{\frac{\pi}{3}}$ CSL is described by the following $U^3(1)$ Chern-Simons theory:
\begin{eqnarray}
\nonumber \mathcal{L}_f &=& \frac{\epsilon^{\mu\nu\lambda}}{4\pi} \sum_{m=-1}^{1} C_m a^m_{\mu} \partial_{\nu} a^m_{\lambda} + \frac{\epsilon^{\mu\nu\lambda}}{2\pi} A^{S^z}_{\mu} \partial_{\nu} \left(\sum_{m=-1}^{1} m \cdot a^m_{\lambda}\right)\\
&=& \frac{\epsilon^{\mu\nu\lambda}}{4\pi} C_{IJ} a_{I\mu}\partial_{\nu}a_{J\lambda}+\frac{\epsilon^{\mu\nu\lambda}}{2\pi} t_I A^{S^z}_{\mu} \partial_{\nu} a_{I\lambda},
\end{eqnarray}
where $A^{S^z}_\mu$ is the gauge potential that couple to the $S^z$ spin density and current, $I,J=1,2,3$, and
\begin{equation}
\begin{array}{lr}
C=\begin{pmatrix}
1&0&0\\
0&1&0\\
0&0&1
\end{pmatrix}, &
t=\begin{pmatrix}
1\\
0\\
-1
\end{pmatrix}
\end{array}.
\end{equation}
The local constraint of the fermions, Eq.~(\ref{constraint}), can be written in a covariant form:
\begin{eqnarray}
\frac{\epsilon^{\mu\nu\lambda}}{2\pi} \sum_{m} \partial_{\nu} a^m_{\lambda} = \sum_{m=-1}^{1} J^{\mu}_m = \bar{J}^{\mu} \equiv \frac{\epsilon^{\mu\nu\lambda}}{2\pi}\partial_\nu \bar{a}_\lambda,~~
\end{eqnarray}
where $\bar{a}_{\mu}$ is a non-dynamical (constant) background field, whose density $\bar{J}^0=(\partial_x \bar{a}_y - \partial_y \bar{a}_x)/2\pi =$ R.H.S. of Eq.~(\ref{constraint}). The constraint can be implemented by introducing an extra $U(1)$ gauge field $b_\mu$ as a Lagrangian multiplier,
\begin{eqnarray}
\nonumber \mathcal{L}_{constraint} &=& \frac{\epsilon^{\mu\nu\lambda}}{2\pi} b_\mu \partial_\nu \left( \sum_{m=-1}^{1} a^m_{\lambda}  - \bar{a}_\lambda \right)\\
&=&\frac{\epsilon^{\mu\nu\lambda}}{2\pi}b_{\mu} \partial_{\nu}\left(\sum_{I}a_{I\lambda}-\bar{a}_{\lambda}\right).
\end{eqnarray}
After integrating out the gauge field $b^\mu$ and $a_{2\mu}$ (or $a^0_\mu$ in the original language), we can obtain the low-energy theory of the $\Phi_{\frac{\pi}{3}}$ CSL state, \cite{pi/3:CS_note}
\begin{eqnarray}
\nonumber \mathcal{L}_{CS} &=& \mathcal{L}_f + \mathcal{L}_{constraint}\\
&=&\frac{\epsilon^{\mu\nu\lambda}}{4\pi}K_{ij}a_{i\mu}\partial_{\nu} a_{j\lambda}+\frac{\epsilon^{\mu\nu\lambda}}{2\pi}q_i A^{S^z}_{\mu}\partial_{\nu}a_{i\lambda},
\end{eqnarray}
with $i,~j=1,2$ (where we relabel $I,J=1,3\rightarrow i, j =1,2$) and the $2\times2$ $K$-matrix and $q$-vector are
\begin{equation}\label{K-matrix:pi/3}
\begin{array}{lr}
K=\begin{pmatrix}
2&1\\
1&2
\end{pmatrix}, &
q=\begin{pmatrix}
1\\
-1
\end{pmatrix}
\end{array}.
\end{equation}
The $\Phi_{\frac{\pi}{3}}$ CSL has the spin quantum Hall conductance (in units of $\frac{1}{2\pi}$),
\begin{eqnarray}
\sigma^{s}_{xy}= q^{T} K^{-1} q = 2.
\end{eqnarray}
Besides, there are two different anyon excitations, both having the self statistical angle $\theta=2\pi/3$. Their mutual braiding statistics is $\theta' = 2\pi/3$. This $\Phi_{\frac{\pi}{3}}$ CSL is similar to the CSL discussed in the footnote [35] of Ref.~\onlinecite{YMLu2012}.

Once we know the effective Chern-Simons theory for the mean-field CSL states, the corresponding edge theory can be obtained from bulk-edge correspondence \cite{Wen1995}. The effective edge theory (assume the edge is along $\hat{x}$-direction) for $\Phi_{\frac{\pi}{3}}$ CSL is
\begin{eqnarray}
\nonumber \mathcal{L}^{\Phi_{\frac{\pi}{3}}}_{edge} = &&  \frac{1}{4\pi} \sum_{i,j=1,2} \left( K_{ij}\partial_t \phi_i \partial_x \phi_j - V_{ij} \partial_x \phi_i \partial_x \phi_j \right) \\
&& + \frac{1}{2\pi} \sum_{i=1,2} q_i \left(A^{S^z}_0 \partial_x \phi_i - A^{S^z}_x \partial_t \phi_i \right),
\end{eqnarray}
where $K$ matrix and $q$ vector are defined in Eq.~(\ref{K-matrix:pi/3}). The $V$ matrix is positive-definite real symmetric which determine the velocities of edge modes and the number of right movers and left movers are determined by the number of positive and negative eigenvalues. We note that in this case, the eigenvalues of $K$ arel {\it positive} and the edges are chiral and stable, at least at the mean-field level here.

The $S^z$ density on the edge are given by the defined bosons in the $\Phi_{\frac{\pi}{3}}$ CSL regime as follows:
\begin{eqnarray}
&& S^z_{\Phi_{\frac{\pi}{3}}}(x) \simeq \sum_{i=1,2} q_i \frac{\partial_x \phi_i (x)}{2\pi},
\end{eqnarray}
and $\phi_1$ carries $S^z$ quantum number $+1$. $\phi_2$ carries $S^z$ quantum number $-1$. 

We can also write down the bosonized expression of the transverse component of the spin on the edge:
\begin{eqnarray}
S^{+}_{\Phi_{\frac{\pi}{3}}}(x) \sim e^{-i (2\phi_1 + \phi_2)} + e^{i(\phi_1 + 2\phi_2)}.
\end{eqnarray}
The edge boson fields in $\Phi_{\frac{\pi}{3}}$ CSL satisfy the Kac-Moody algebra:
\begin{eqnarray}
\left[ \phi_j (x) , \partial_x \phi_k (y) \right] = i 2\pi (K^{-1})_{jk} \delta(x-y).
\end{eqnarray}
At the mean-field level, because of the gapless edge excitations in the CSL phase, the transverse spin components should show power law $\la S^{+}(x,t) S^{-}(0,t) \ra \sim |x|^{-p}$, where $p$ is some number determined by the details of the matrix $V$ in $\Phi_{\frac{\pi}{3}}$ CSL.

\subsubsection{Effective Chern-Simons theory for mean-field $\Phi_{\frac{2\pi}{3}}$ CSL}\label{Subsubsec:CS42pi/3}
Now the three flavors of fermions fill the lowest bands with Chern numbers $C_{\pm1}=C_0=+2$.  The fermion band structure in this $\Phi_{\frac{2\pi}{3}}$ CSL is described by the following $U^6(1)$ Chern-Simons theory:
\begin{eqnarray}
\nonumber \mathcal{L}_f &=& \frac{\epsilon^{\mu\nu\lambda}}{4\pi} \sum_{l,k=1,2}\sum_{m=\pm1,0} \left[ a_{l,\mu}^{m} \partial_\nu a_{k,\lambda}^m + 2 A^{S^z}_{\mu} m \partial_{\nu} a^m_{l,\lambda} \right]\\
&=& \frac{\epsilon^{\mu\nu\lambda}}{4\pi} C_{IJ} a_{I\mu}\partial_{\nu}a_{J\lambda}+\frac{\epsilon^{\mu\nu\lambda}}{2\pi} t_I A^{S^z}_{\mu} \partial_{\nu} a_{I\lambda},
\end{eqnarray}
where $A^{S^z}_\mu$ is the gauge potential that couple to the $S^z$ spin density and current, $I,J=1,2,...,6$, and
\begin{equation}
\begin{array}{lr}
C=(\mathbbm{1})_{6\times6}, &
t=\begin{pmatrix}
1&1&0&0&-1&-1
\end{pmatrix}^{T}
\end{array},
\end{equation}
where $(\mathbbm{1})_{n\times n}$ represents $n\times n$ identity matrix. The local constraint now can be written as,
\begin{eqnarray}
\frac{\epsilon^{\mu\nu\lambda}}{2\pi}\sum_{I}\partial_{\nu}a_{I\lambda}=\sum_{I=1}^{6}J_I^\mu=\bar{J}^\mu\equiv \frac{\epsilon^{\mu\nu\lambda}}{2\pi}\partial_\nu \bar{a}_\lambda.
\end{eqnarray}
The constraint can be implemented by introducing an extra $U(1)$ gauge field $b_\mu$ as a Lagrangian multiplier,
\begin{eqnarray}
\mathcal{L}_{constraint}=\frac{\epsilon^{\mu\nu\lambda}}{2\pi}b_{\mu}\partial_\nu\left(\sum_{I=1}^{6}a_{I\lambda} -\bar{a}_{\lambda}\right).
\end{eqnarray}

After integrating out the gauge fields $b^\mu$ and the gauge field, $a_{3,\mu}$ (or $a^0_{1\mu}$), we can obtain the effective Chern-Simons theory for the $\Phi_{\frac{2\pi}{3}}$ CSL: \cite{2pi/3:CS_note}
 \begin{eqnarray}
 \nonumber \mathcal{L}_{CS}&=&\mathcal{L}_f + \mathcal{L}_{constraint}\\
 &=& \frac{\epsilon^{\mu\nu\lambda}}{4\pi} K'_{ij} a_{i\mu}\partial_\nu a_{j\lambda} + \frac{\epsilon^{\mu\nu\lambda}}{2\pi} q'_i A^{S^z}_{\mu}\partial_\nu a_{i\lambda},
 \end{eqnarray}
 with $i,j=1,2,..,5$ (where we relabel $I,J = 1,2,4,5,6 \rightarrow i,j=1,2,3,4,5$) and the $K'$-matrix and $q'$-vector are
 \begin{equation}\label{K-matrix:2pi/3}
 \begin{array}{lr}
 K'=\begin{pmatrix}
2&1&1&1&1\\
1&2&1&1&1\\
1&1&2&1&1\\
1&1&1&2&1\\
1&1&1&1&2
 \end{pmatrix},&
 q'= \begin{pmatrix}
 1&1&&0&-1&-1
 \end{pmatrix}^{T}
 \end{array}.
 \end{equation}
 Therefore, the $\Phi_{\frac{2\pi}{3}}$ CSL has the spin quantum Hall conductance,
 \begin{eqnarray}
 \sigma^{s}_{xy} = q'^T K'^{-1} q' = 4.
 \end{eqnarray}
 There are five anyon excitations, each having the statistical angle $\theta=5\pi/6$. However, the mutual braiding statistics between each two is $\theta'=4\pi/3$.

The effective edge theory for $\Phi_{\frac{2\pi}{3}}$ CSL is
\begin{eqnarray}
\nonumber \mathcal{L}^{\Phi_{\frac{2\pi}{3}}}_{edge} = &&  \frac{1}{4\pi} \sum_{i,j=1,..,5} \left( K'_{ij}\partial_t \phi'_i \partial_x \phi'_j - V'_{ij} \partial_x \phi'_i \partial_x \phi'_j \right) \\
&& + \frac{1}{2\pi} \sum_{i=1}^{5} q'_i \left(A^{S^z}_0 \partial_x \phi'_i - A^{S^z}_x \partial_t \phi'_i \right),
\end{eqnarray}
where $K'$ matrix and $q'$ vector are defined in Eq.~(\ref{K-matrix:2pi/3}). $V'$ matrix is positive-definite real symmetric and the eigenvalues of $K'$ are all {\it positive} and the edges are chiral and stable, at least at the mean-field level here.

The $S^z$ density on the edge in the $\Phi_{\frac{2\pi}{3}}$ CSL regime is
\begin{eqnarray}
S^z_{\Phi_{\frac{2\pi}{3}}}(x) \simeq \sum_{i=1,2,4,5}  q'_i \frac{\partial_x \phi'_i(x)}{2\pi},
\end{eqnarray}
where $\phi'_1,$ and $\phi'_2$ carry $S^z$ quantum number $+1$. $\phi'_{4}$ and $\phi'_{5}$ carry $S^z$ quantum number $-1$. The remaining $\phi'_3$ carries $S^z$ quantum number $0$.

The bosonized expression of the transverse component of the spin on the edge is
\begin{eqnarray}
\nonumber && S^{+}_{\Phi_{\frac{2\pi}{3}}}(x)\sim e^{-i(2\phi'_1+\phi'_2+\phi'_4+\phi'_5)}+e^{i(\phi'_1+\phi'_2+2\phi'_4+\phi'_5)}+\\
&&\hspace{1.8cm} +e^{-i(\phi'_2-\phi'_3)}+e^{-i(\phi'_3-\phi'_5)}.
\end{eqnarray}
The edge boson fields satisfy the Kac-Moody algebra:
\begin{eqnarray}
\left[ \phi'_j (x) , \partial_x \phi'_k (y) \right] = i 2\pi (K'^{-1})_{jk} \delta(x-y).
\end{eqnarray}
At the mean-field level, the transverse spin components should show power law $\la S^{+}(x,t) S^{-}(0,t) \ra \sim |x|^{-p'}$, where $p'$ is some number determined by the details of the $V'$ matrix in $\Phi_{\frac{2\pi}{3}}$ CSL.

\section{Discussion}\label{Sec:Discussion}
We study the SU(3) model with AFM three-site ring exchanges. The site-factorized state studies over the magnetic ordered states show two different three-sublattice-ordered states, AFQ and $120^o$ AFM, along the $K_3/J$ axis. We later use slave-fermion trial states to study the non-magnetic states. We find that there are three competing trial states--trimer, $\Phi_{\frac{\pi}{3}}$ CSL, and $\Phi_{\frac{2\pi}{3}}$ CSL. At the mean-field level, the $\Phi_{\frac{\pi}{3}}$ CSL has Chern number $C_{\Phi_{\frac{\pi}{3}}}=1$ with a finite spin quantum Hall conductance $\sigma_{xy}=2$. There are two different anyon excitations, both having statistical angle $\theta=2\pi/3$, but their mutual braiding statistics is $\theta' = 2\pi/3$;Interestingly the $\Phi_{\frac{2\pi}{3}}$ CSL has Chern number $C_{\Phi_{\frac{2\pi}{3}}}=2$ with finite spin quantum Hall conductance $\sigma_{xy} =4$. There are five different anyon excitations, each having statistical angle $\theta = 5\pi/6$, but the mutual braiding statistics of each two is $\theta' =4\pi/3$.

Within the scope of this work, it is not possible to compare directly the energetics between the magnetic ordered states and the non-magnetic states since we used two different methods to explore the phases in this model. The only wave function of the non-magnetic state that we know is the trimer state whose energy can be lowered than that of the $120^o$ AFM in a small parameter window as discussed in the last paragraph of Sec.~\ref{Subsec:MFcalc}. Since at the mean field, the uniform-flux CSL have lower energies than that of the trimer in a large parameter regime, the comparison above gives us the possibility that the uniform-flux CSL will be realized after performing the Gutzwiller projection on the mean-field states. 

It is possible to perform linear flavor wave theory  (LFWT) as illustrated in Ref.~\onlinecite{Bauer2012} to study the effects of quantum fluctuations on the magnetic ordered states. However, the quantum fluctuations corrections in this model actually lower the energies of the magnetic ordered states as shown in the note \onlinecite{LFWT:AFQ_note} for the AFQ. Such qualitative arguments about the quantum fluctuations destroying the magnetic order hence can not be directly applied here and more qualitative analysis such as Variational Monte Carlo (VMC), DMRG, and iPEPs are needed. The parameter regime in which the uniform-flux CSL are more likely to be detected by the numerics can be seen as follows. According to the VMC results in Ref. \onlinecite{Bieri2012}, the interesting $d_x + i d_y$ CSL is realized in a small window around $K_3/J \sim 1$ between the two magnetic ordered phase--AFQ and $120^o$ AFM. Intuitive, in our mean field picture, since the uniform-flux CSL have much lower energies than those of the gapless QSL, which lead to the interesting $d_x + i d_y$ CSL via plausible pairing mechanisms, the small window of the $d_x + i d_y$ CSL is likely replaced by that of the uniform-flux CSL, especially the $\Phi_{\frac{\pi}{3}}$ CSL since $\Phi_{\frac{2\pi}{3}}$ CSL is deep inside the $120^o$ AFM regime.

In the present work, we ignore the gauge fluctuations in the mean-field slave-fermion analysis. The gauge fluctuations around our mean-field results can be described by the compact U(1) lattice gauge theory in (2+1)D, which brings out the confinement issue due to the proliferation of the topological defects. Indeed, in (2+1)D, if the partons (spinons) in the U(1) slave-fermion theory are gapped without breaking Time-Reversal symmetry (such as $\pi$-flux) and the filled bands are {\it trivial}, a.k.a. Chern number $C=0$, the topological defects along the space-time known as the instantons or Dirac monopoles are proliferating to induce the instabilities that lead the gapped QSL to some order states which break translational invariance. \cite{Polyakov1977} In sharp contrast, if the filled parton bands are topological as in the present uniform-flux CSL, the effective theory of the gauge fluctuations is described by the U(1) lattice gauge theory with a Chern-Simons term. The Chern-Simons term has dramatic effects on the behavior of the Dirac monopoles. The Dirac monopoles become charged and linearly confined due to the presence of a Chern-Simons term \cite{Pisarski1986, Affleck1989, Fradkin1991, Lee1992, Diamantini1993}, and the uniform-flux CSL can remain stable against gauge fluctuations. 

The topological order in the CSL phases can be partially detected using entanglement spectrum and the topological entanglement entropy (Renyi entropy or von Neumann entropy). \cite{Li_Haldane, Grover2012, Cincio2012, Zhang2012, Grover_TEE_review} The Chern number of the CSL phases can be possibly extracted using Exact Diagonalization.\cite{Wang2012} Furthermore, recently it is also suggested that the quasi-particle self and mutual braiding statistics can be extracted from the ground states with minimum entropies using VMC. \cite{Zhang2012}

For a possible future direction of the present theoretical model, even though it may be difficult to detect the interesting uniform-flux CSL in the cold atom experiments, since closer to the metal-Mott insulator phase transition the higher order terms such as four-site ring exchange terms also needs consideration, the present model can serve as a theoretical starting point for a lattice model realization of the interesting $U_{S^z}(1)$ symmetry protected topological phase (SPT) in the featureless spin-1 frustrated magnet. \cite{YMLu2012} Following the logic detailed in Ref.~\onlinecite{YMLu2012}, the $U_{S^z}(1)$ SPT can be possibly realized in the present model by introducing a Time-Reveral breaking (TRB) and $S^z$-conserving two-site exchange term. The possible Hamiltonian is
\begin{eqnarray}
\nonumber H_{U(1)-SPT} = J \sum_{\linepic} P_{12} &+& K_{3} \sum_{\triangpic}  \left[ P_{123} +\Hc \right]+ \\
&+&  K_{TRB}\sum_{\linepic} \sum_{\alpha \not= \beta} P^{\alpha \beta}_{12},
\end{eqnarray}
where $\alpha,~\beta = x,~y,$ and $z$. The introduced TRB and $S^z$-conserving $P^{\alpha \beta}_{12}$ in the slave-fermion representation is expressed as $
P^{\alpha \beta}_{12} = e^{i(\theta_{12}^{\alpha} - \theta_{12}^\beta)} f^{\beta \dagger}_{1}f^{\alpha}_1 f^{\alpha \dagger}_{2} f^{\beta}_{2},$ which contains the complex phases as $e^{i (\theta_{12}^\alpha - \theta_{12}^\beta)}$ with $\theta_{12}^{x} = \theta_{12}^{y} = - \theta_{12}^{z}$ that realizes the $U_{S^z}(1)$ symmetry of the rotation about $S^z$-axis. This $K_{TRB}$ term adds another axis to the parameter spaces in the present model and whether or not the $U_{S^z}(1)$ SPT can be energetically favored is interesting by its own.

\acknowledgments
We would like to thank Olexei I. Motrunich, Ling Wang, Dima Pesin, Jason Alicea, and Shu-Ping Lee for helpful discussion. We also acknowledge the support by the National Science Foundation through grant No. DMR-1004545.
\bibliography{biblio4SU3}

\begin{thebibliography}{53}
\bibitem [{ene()}]{energetics_note}%
  \BibitemOpen
  \href@noop {} {}\bibinfo {note} {We also numerically check the paired-fermion
  states (superconducting states), which arise from the pairing instabilities
  of either zero-flux or $\pi$-flux gapless spin liquid states, with pairing
  symmetry of $s$-wave, $p_x+i p_y$, $d_x+id_y$, $f$-wave, etc. We still find
  the optimal trial states in the AFM $K_3$ regime are trimer,
  $\Phi_{\frac{\pi}{3}}$ CSL, and $\Phi_{\frac{2\pi}{3}}$ CSL.}\BibitemShut
  {Stop}%
\bibitem [{pi/()}]{pi/3:CS_note}%
  \BibitemOpen
  \href@noop {} {}\bibinfo {note} {The gauge fixing can be easily done by
  observing that in the continuum limit the constant in the local constraint is
  not important and can be dropped. Therefore, the gauge fixing can be achieved
  by setting $\sum_{m=-1}^{1} J^{\mu}_m=0$, which gives $a^{0}_\mu =
  -(a^{1}_\mu + a^{-1}_\mu )$. With the substitution of $a^{0}_\mu$ by $ - (
  a^{1}_\mu + a^{-1}_\mu)$, we can obtain effective Chern-Simons
  theory.}\BibitemShut {Stop}%
\bibitem [{2pi()}]{2pi/3:CS_note}%
  \BibitemOpen
  \href@noop {} {}\bibinfo {note} {The gauge fixing can be achieved by setting
  $\sum_{m=-1}^{1} \sum_{a=1,2} J^{\mu}_{a,m} = 0$, which gives $a^{0}_{1\mu} =
  -(a^{+1}_{1\mu} + a^{+1}_{2\mu} + a^{-1}_{1\mu} + a^{-1}_{2\mu} +a^0_{2\mu}
  )$. In the new labeling, we replace $a_{3\mu}$ by
  $-(a_{1\mu}+a_{2\mu}+a_{4\mu}+a_{5\mu} + a_{6\mu})$, and we can obtain the
  effective Chern-Simons theory.}\BibitemShut {Stop}%
\end{thebibliography}
\end{document}